# Oxygen Electromigration and Energy Band Reconstruction Induced by Electrolyte Field Effect at Oxide Interfaces

S. W. Zeng[1,2*], X. M. Yin[2,3,4], T. S. Herng[5], K. Han[1,2], Z. Huang[1,2], L. C. Zhang[1,2], C. J. Li[1,5], W. X. Zhou[1,2], D. Y. Wan[1,2], P. Yang[3], J. Ding[1,5], A. T. S. Wee[2,6], J. M. D. Coey[1,7], T. Venkatesan[1,2,5,8,9], A. Rusydi[1,2,3], A. Ariando[1,2,9*]

[1]NUSNNI-NanoCore, National University of Singapore, Singapore 117411, Singapore
[2]Department of Physics, National University of Singapore, Singapore 117542, Singapore
[3]Singapore Synchrotron Light Source (SSLS), National University of Singapore, 5 Research Link, Singapore 117603, Singapore
[4]SZU-NUS Collaborative Innovation Center for Optoelectronic Science & Technology, Key Laboratory of Optoelectronic Devices and Systems of Ministry of Education and Guangdong Province,
College of Optoelectronic Engineering, Shenzhen University, Shenzhen 518060, China
[5]Department of Materials Science and Engineering, National University of Singapore, Singapore 117576, Singapore
[6]Centre for Advanced 2D Materials and Graphene Research, National University of Singapore, Singapore 117546, Singapore
[7]School of Physics and CRANN, Trinity College, Dublin 2, Ireland
[8]Department of Electrical and Computer Engineering, National University of Singapore, Singapore 117576, Singapore
[9]National University of Singapore Graduate School for Integrative Sciences and Engineering (NGS), 28 Medical Drive, Singapore 117456, Singapore
*To whom correspondence should be addressed.
E-mail: phyzen@nus.edu.sg, ariando@nus.edu.sg

**Abstract**

Electrolyte gating is a powerful means for tuning the carrier density and exploring the resultant modulation of novel properties on solid surfaces. However, the mechanism, especially its effect on the oxygen migration and electrostatic charging at the oxide heterostructures, is still unclear. Here we explore the electrolyte gating on oxygen-deficient interfaces between $SrTiO_3$ (STO) crystals and $LaAlO_3$ (LAO) overlayer through the measurements of electrical transport, X-ray absorption spectroscopy (XAS) and photoluminescence (PL) spectra. We found that oxygen vacancies ($O_{vac}$) were filled selectively and irreversibly after gating due to oxygen electromigration at the amorphous LAO/STO interface, resulting in a reconstruction of its interfacial band structure. Because of the filling of $O_{vac}$, the amorphous interface also showed an enhanced electron mobility and quantum oscillation of the conductance. Further, the filling effect could be controlled by the degree of the crystallinity of the LAO overlayer by varying the growth temperatures. Our results reveal the different effects induced by electrolyte gating, providing further clues to understand the mechanism of electrolyte gating on buried interfaces and also opening a new avenue for constructing high-mobility oxide interfaces.

Electric field effect doping, which is the foundation of modern semiconductor electronics, has been a flexible and powerful technique for tuning the carrier density and the resultant transport properties of a material [1]. More recently, the electric field provided by electronic double layer transistor (EDLT) with ionic liquids (ILs) or polymer electrolytes as the gate dielectrics has enabled the accumulation of a large amount of charge carriers up to $\sim 10^{15}/cm^2$ [2], and consequently novel phase transitions in a variety of materials have been induced [3-16]. In an EDLT, the induced charged layer on the sample surface may be a direct result of electrostatic carrier injection [3-11,13,14,17], migration of oxygen [15,16,18-26] and hydrogen [16,27-29], and/or structural changes [30-33]. Specifically, for the oxygen-free two-dimensional (2D) layered materials, the dominant driving mechanism has been shown to be electrostatic charging, and thus allowing the study of ion gating of clean 2D superconductors [4,6-11]. In complex oxides, however, oxygen migration is possible due to the electrochemical reaction on the sample surface and the large electric field as the IL is directly applied on the oxide surface [15,18,19]. Through insertion of a chemically inert layer between the IL and conducting channel, the oxides can be protected against electrochemical reactions, leading to an enhancement of the carrier mobility [34-36]. However, the gating mechanism, especially the effect of oxygen electromigration on buried oxides under the protective overlayer and the related influence on the band structure is still unclear [34-38].

STO-based oxide heterostructures such as the interface with LAO overlayer are of a great interest [39-45]. Tunability of electrical properties at the LAO/STO interface could be realized using various methods such as field effect [42], metal capping [46], surface adsorbates [45] and defect control [47], suggesting its potential for oxide-based electronics [48, 49]. Conducting interfaces can be even constructed at room temperature by depositing amorphous films on crystalline STO, which have the advantage of being compatible with established semiconductor fabrication processes [50-52]. Moreover, high carrier mobility at amorphous LAO/STO (*a*-LAO/STO) interface is achieved by charge-transfer-induced modulation doping, through inserting a $La_{1-x}Sr_xMnO_3$ layer as the potential barrier, suggesting the importance of interfacial band structure for charge transport [53]. In this paper, we demonstrate IL gating induced oxygen electromigration on LAO/STO interfaces and the resultant change of band structure due to the selective filling of interfacial $O_{vac}$.

The interfaces were obtained by depositing LAO on STO substrates using a pulsed laser deposition (PLD) system. The STO are (100)-oriented unless otherwise noted. The deposition temperature ($T_d$) of *a*-LAO is 25 °C and oxygen partial pressure is $P_{O2}$ = 2 × $10^{-6}$ Torr unless otherwise noted. Before deposition, patterns with a Hall bar and a lateral gate pad were fabricated on STO by photolithography (Fig. S1(a)) [54]. The gate electrode was formed by covering silver paint on the lateral gate pad. The wire connection for transport measurement was done by Al ultrasonic wire bonding. A small drop of the IL, N,N-diethyl-N-methyl-N-(2-methoxyethyl)ammonium bis(trifluoromethyl sulphonyl)imide (DEME-TFSI), covered both the channel and gate electrode. The transport measurements were made in Quantum Design Physical Property Measurement System (PPMS), using the built-in Source Measure Units, Keithley 2400 Sourcemeters and 2002 Multimeters. For the XAS and PL measurements, unpatterned large-area samples were prepared (Fig. S1(b)) [54]. The PL measurements were performed under excitation with a He-Cd laser line (325 nm) and the spectra were collected by iHR 550 spectrometer. The XAS measurements were performed using the Surface, Interface and Nanostructure Science beamline in Singapore Synchrotron Light Source. The incident X-ray is normal to the sample surfaces.

Figure 1(a) shows the resistance as a function of $V_G$ for a device over ten scan cycles (see Fig. S1(c) for the schematic of an EDLT device) [54]. The device could be switched between low and high-resistance states with hysteresis and irreversibility. The irreversible behaviour is more obvious in the first cycle. After the 3$^{rd}$ cycle, the scanning is reversible, even though a slight hysteresis is still observed (Inset of Fig. 1(a)). For comparison, Figure 1(b) shows the resistance of an oxygen-annealed crystalline LAO/STO device in which O$_{vac}$ are absent [52]. A reversible and non-hysteretic metal-insulator transition is observed, indicating pure electrostatic charging and depletion [36]. Moreover, the $a$-LAO/STO remains insulating even when the $V_G$ is reversed back to 0 V. These results indicate that electrostatic effect is not the only mechanism of IL gating on the $a$-LAO/STO interface.

Figures 2(a)-(c) shows sheet resistance $R_s$, carrier density $n_s$ and Hall mobility $\mu_H$ as a function of temperature $T$ for samples before and after gating. The measurements before gating were performed on the pristine device before any IL was applied. The ones after gating were performed after the devices were gated for different cycles (1 and 10 cycles) and stopped at different $V_G$ (-2, 0 and 2 V), by scanning $V_G$ between 2 and -2 V, and then

subsequently removing the IL. One can see that all samples are metallic due to the creation of $O_{vac}$ in STO during the LAO deposition [50-52]. However, the samples show higher $R_s$ at high $T$ and lower $R_s$ at low $T$ after gating, suggesting improved metallic conduction. Moreover, the samples after gating show similar behavior of $R_s$-$T$ curves, even though the samples stopping at -2 V show slightly lower $R_s$ at low $T$, compared to those stopping at 0 and 2 V.

The sample before gating shows $n_s$ of ~$1.21 \times 10^{14}$ cm$^{-2}$ at 300 K and ~$2.71 \times 10^{13}$ cm$^{-2}$ at 4 K. The $n_s$ decreases with decreasing $T$ below ~150 K, suggesting carrier freeze out which is characterized by an activation energy $\varepsilon$ of ~5 meV (fitted by $n_s \propto e^{(-\varepsilon/k_B T)}$) [55], and therefore, the presence of localized electrons. In contrast, $n_s$ of the samples after gating is nearly $T$ independent and show a much lower value of ~$2 \times 10^{13}$ cm$^{-2}$, suggesting that part of $O_{vac}$ was filled. Due to the lower $R_s$ and $n_s$, the samples after gating show higher $\mu_H$ at low $T$ (Fig. 2(c)). For example, the sample gated for 10 cycles and stopping at -2 V shows $\mu_H$ of ~1610 cm$^2$/Vs at 4 K, higher than that of the sample before gating (~550 cm$^2$/Vs). The highest $\mu_H$ of ~3580 cm$^2$/Vs at 3 K is obtained on the 15-nm $a$-LAO/STO interface, and consequently, quantum oscillations of the magnetoresistance can be observed (Fig. S2) [54]. Figure 2(d) shows the $R_s$ and $n_s$ at 300 K for samples before and after gating (10 cycles and stop at -2 V) with different $a$-LAO thickness. We also performed the gating on $a$-LAO/STO interfaces with different STO orientations, and interfaces between STO substrates and a series of other amorphous materials (Figs. S3 and S4) [54]. All samples after gating show one order of magnitude higher $R_s$ and lower $n_s$, further suggesting the universal decrease in $O_{vac}$.

Figures 2(e) and 2(f) show the $R_s$ and $n_s$ for samples before and after application of one single $V_G$ pulse for different durations. The $R_s$ and $n_s$ of samples after $V_G$ = +2 V pulses are almost unchanged, compared with the sample before gating. In contrast, the samples after -2 V pulses show one order of magnitude higher $R_s$ and lower $n_s$. An example of resistance as a function of time for one single 60-second gate pulse is shown in Fig. 2(g). These results indicate that oxygen filling can be induced only after application of negative $V_G$ which causes the samples to be insulating states. Even with a -2 V pulse duration of only 1 second, a large change of $R_s$ and $n_s$ was still observed, suggesting that the filling of $O_{vac}$ is a transient process.

Figure 3(a) shows the room-temperature Ti $L_{2,3}$-edge XAS spectra of 2.5-nm $a$-LAO/STO interfaces. In oxygen-deficient STO, charge transfer from $O_{vac}$ to Ti atoms, causes the change of Ti oxidation from $Ti^{4+}$ to $Ti^{3+}$, and therefore, the corresponding vacancy concentration can be derived. As indicated by the arrow, the intensity of 458 eV emission, which comes from the $Ti^{3+}$ $L_3$ edge [56, 57], decreases after gating. This means that the amount of $Ti^{3+}$, and therefore the number of $O_{vac}$ decrease after gating. The decrease in $O_{vac}$ can also be proved from the O $K$-edge spectra in which the peak intensity arising from Ti-O $pd$ hybridization decreases after gating (Fig. S5) [54,56]. Figure 3(b) shows the room-temperature PL spectra. One can see a broad emission centred at ~430 nm, which is similar to that observed in $Ar^+$-irradiated STO and is caused by the radiative process between the doped conduction electrons and the in-gap state [58], indicating that the emission here originates from the oxygen-deficient STO below $a$-LAO [52, 58]. The PL intensity decreases after gating further indicates the decrease in $O_{vac}$ concentration. Furthermore, the *in-situ* PL showed that with decreasing $V_G$, the intensity decreases continuously, directly indicating oxygen electromigration and the resultant decrease in $O_{vac}$ during the gating process (Fig. S5) [54].

Figures 4(a) and 4(b) show the $n_s$ and $R_s$ at 300 K before and after gating for LAO/STO interfaces with $T_d$ from 25 to 780 °C. All samples were deposited at the same $P_{O2} = 1\times10^{-4}$ Torr without post oxygen annealing, and therefore, $O_{vac}$ which dominate the charge carriers are present [52]. The pristine interfaces show similar $n_s$ of $\sim 1\times10^{14}$ cm$^{-2}$ and $R_s$ of $\sim$10000 $\Omega/\square$. After gating, the samples show general decrease in $n_s$ and increase in $R_s$ with decreasing $T_d$ below 700 °C. For $T_d$ above 500 °C at which LAO is crystalline [59], samples grown at 780 and 700 °C show almost unchanged $n_s$ and $R_s$ after gating, and samples grown at relative low $T_d$ of 600 and 500 °C show slight decrease in $n_s$ and increase in $R_s$. For $T_d$ below 500 °C at which LAO is amorphous [59], the changes of $n_s$ and $R_s$ after gating are more obvious with decreasing $T_d$, and one order of magnitude change is observed at 25 °C. These suggest that more $O_{vac}$ were filled at interfaces with lower $T_d$, and thus, the filling effect could be controlled by $T_d$.

The $n_s$ as a function of $T$ for LAO/STO interfaces with different $T_d$ were also measured (Fig. S6) [54]. The $\varepsilon$ before and after gating for each sample is shown in Fig. 4(c). Before gating, $n_s$ of all samples show similar carrier freeze out behaviour and the $\varepsilon$ are similar ranging from 5 to 7 meV. In contrast, the samples show less carrier freeze out effect after gating with decreasing $T_d$, and for $T_d$ =25 °C, the $n_s$ is independence of $T$.

Correspondingly, the $\varepsilon$ decreases from ~6 to 0 meV as the $T_d$ decreases from 700 to 25 °C.

Figure 5 shows a schematic band diagram of the LAO/STO interface [60-63]. The carriers are thermally activated from $O_{vac}$ donor level ($E_{Ov}$) to conduction band ($E_C$) and accumulate at the interface at high $T$. As $T$ decreases, more and more free electrons freeze and are trapped at $E_{Ov}$, and therefore, a carrier freeze-out effect is observed in the samples before gating (Fig. 5(a) and Fig. S6). After gating, the carrier behaviour is the same as that before gating for the interface grown at 700 °C, since the filling is negligible and $E_{Ov}$ is unchanged (Fig. 5(a)). For the interface grown at 300 °C, part of $O_{vac}$ are irreversibly filled and $E_{Ov}$ move closer to $E_C$ after gating (Fig. 5(b)), and thus the $n_s$ decreases and shows less carrier freeze-out effect (Fig. S6). For the interface grown at 25 °C, more $O_{vac}$ are filled and $E_{Ov}$ moves to the same level as $E_C$ after gating (Fig. 5(c)). The free electrons are present at $E_C$ rather than trapped at $E_{Ov}$ as $T$ decreases, and thus, the carrier freeze-out effect is absent (Fig. S6). Since the $n_s$ decreases after filling, the Fermi level ($E_F$) shifts down and the bands at the STO side show less bending accordingly. Therefore, the residual $E_{Ov}$ after filling are much closer to $E_C$ with decreasing $T_d$ and the decrease in $\varepsilon$ is observed, indicating that the residual $E_{Ov}$ are reconstructed and the reconstruction effect strongly depends on $T_d$. After filling, the residual electrons can be induced and depleted reversibly (Inset of Fig. 1(a)). Moreover, after removing the IL, the samples show similar behaviour of $R_s$ and $n_s$, regardless of the $V_G$ at which the scanning stops (Fig. 2(a) and 2(b)). This is different from the observation in $VO_2$, the sample maintained its state at the $V_G$ where the scanning stopped [15]. These probably suggest that after partial filling of $O_{vac}$, the gating effect on the residual carriers is mainly a result of electrostatic charging and depleting.

The deposited overlayer appears to play an important role on the oxygen filling. It has been found that the filling of $O_{vac}$ is not obvious when the LAO layer is crystalline (Fig. 4). We also found that gating performed in the PPMS chamber where oxygen is absent and in air show similar results, suggesting that the filling is not dependent on the oxygen gas environment. Therefore, we deduce that it is the oxygen in the amorphous overlayer that migrates into STO to fill the vacancies, under the influence of electric field (Fig. S1(d)) [54]. Note that the occurrence of $O_{vac}$ in LAO can be compensated by a charged interface [47]. Since the filling causes the change of Ti valence states from $Ti^{3+}$ to $Ti^{4+}$, the redox reaction during gating process is expected. It has been reported that the surfaces show considerable damage after the samples (single crystalline $TiO_2$, STO and $La_{0.5}Sr_{0.5}CoO_{3-x}$) were irreversibly gated from insulator to metal, due to the electrochemical redox reaction on the sample surfaces [18, 23, 64]. In the present results, the atomic force microscopy images of the sample surfaces before and after gating are essentially unchanged (Fig. S7) [54]. Moreover, the leakage current is negligibly small, on the order of $10^{-9}$ A. These suggest that the redox reaction only occurs on the STO near the interface, not on the LAO surface or within the electrolyte.

In conclusion, we have demonstrated the electrolyte gating-induced oxygen migration and the resultant filling of $O_{vac}$ at LAO/STO interfaces. The filling effect caused the reconstruction of interfacial band structures, which could be controlled by the overlayer deposition temperature, suggesting a method for tuning band structures and oxidation states on buried oxides. Since the filling occurs on the substrate surface, our findings would also be useful for future exploration of electrolyte gating on various emergent oxide interfaces [65, 66] and interfaces based on oxide substrates such as superconducting FeSe/STO [67]. Moreover, enhanced carrier mobility and quantum magnetoresistance oscillations are achieved as a result of gating-induced oxygen filling, opening an avenue to construct high-mobility oxide heterostructures which are synthesized at room temperature.

## Acknowledgments

This work is supported by the National University of Singapore (NUS) Academic Research Fund (AcRF Tier 1 Grant No. R-144-000-346-112 and R-144-000-364-112) and the Singapore National Research Foundation (NRF) under the Competitive Research Programs (CRP Award No. NRF-CRP15-2015-01, CRP Award No. NRF-CRP10-2012-02 and CRP Award No. NRF-CRP 8-2011-06). P. Yang is supported from SSLS via NUS Core Support C-380-003-003-001.

**Figure legends**

**Figure 1.** (a) Resistance $R=V_{SD}/I_{SD}$, where $V_{SD}$ is the source-drain voltage and $I_{SD}$ is the source-drain current, as a function of $V_G$ for a 2.5-nm $a$-LAO/STO device. Inset of (a) is the same scanning from 4th to 10th cycles. At each cycle, the scanning of $V_G$ starts from 0 to +2 to -2 V and stops at 0 V. (b) Resistance as a function of $V_G$ for an oxygen-annealed crystalline LAO/STO device. The $V_G$ is swept at a speed of 25 mV/s and $V_{SD}$=3 V is applied.

**Figure 2.** (a) The $R_s$-$T$ curves of 2.5-nm $a$-LAO/STO samples before and after gating. (b) The $n_s$ and (c) $\mu_H$ as a function of temperature. (d) $R_s$ and $n_s$ for samples with $a$-LAO at different thickness. (e) $R_s$ and (f) $n_s$ for four samples before and after application of one single $V_G$ pulse for different duration, $V_G$ = -2 and +2 V for 1, 10, 60, 300 seconds. For convenient comparison, the data before gating is labelled at 1 second. (g) Resistance as a function of time for one single 60-second $V_G$ pulse at -2 and +2 V.

**Figure 3.** (a) Ti $L_{2,3}$-edge XAS and (b) PL spectra for 2.5-nm $a$-LAO/STO interfaces before and after gating. The black arrow indicates the emission coming from $Ti^{3+}$.

**Figure 4.** (a) The $n_s$ and (b) $R_s$ as a function of $T_d$ for samples before and after application of a negative $V_G$ pulse for duration of 2 min. (c) The $\varepsilon$ as a function of $T_d$. The dot lines are the guides for eyes.

**Figure 5.** Schematic band diagrams of the LAO/STO interfaces after gating for $T_d$ of (a) 700 °C, (b) 300 °C and (c) 25 °C. For the interfaces before gating, the band diagrams are the same as shown in (a), since the $R_s$ and $n_s$ are the same as those after gating at 700 °C. The single arrow in (a) denotes the transition of electrons from $E_{Ov}$ to $E_C$ as a result of thermal activation. The short lines in the band gap denotes the $E_{Ov}$ and the circles denote the holes at $E_{Ov}$ after the electron transition to $E_C$. The $\varepsilon$ after gating are also shown.

**Figures**

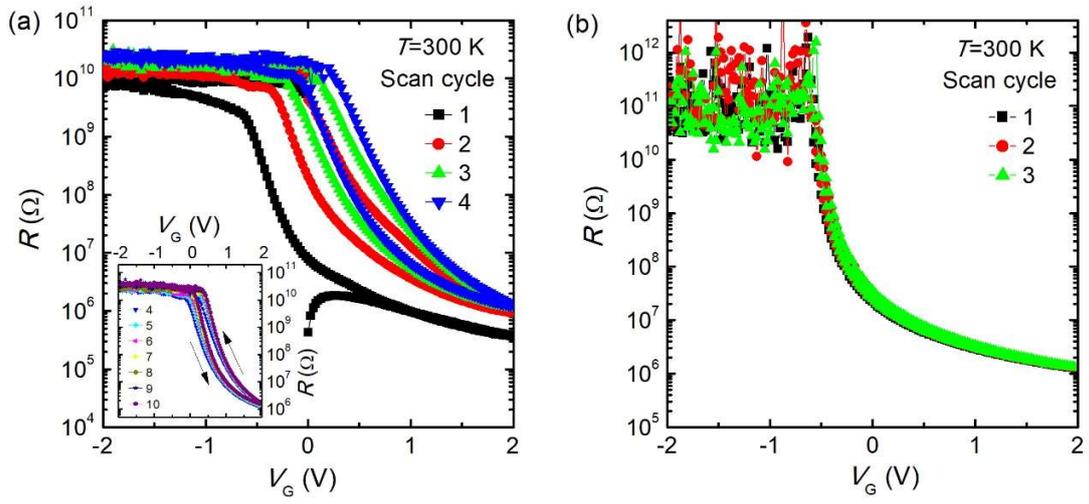

Figure 1

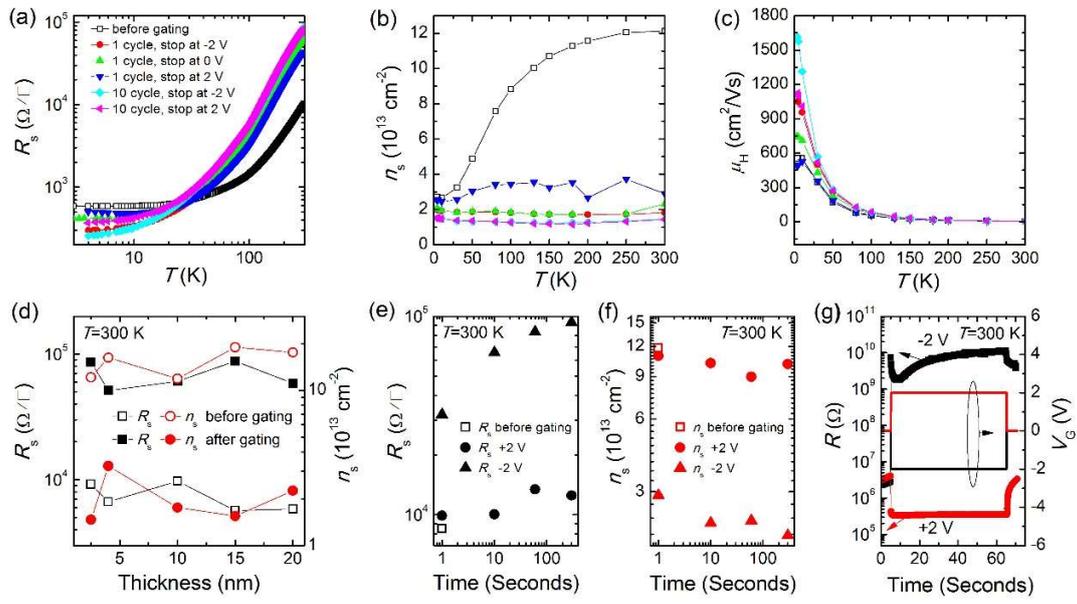

Figure 2

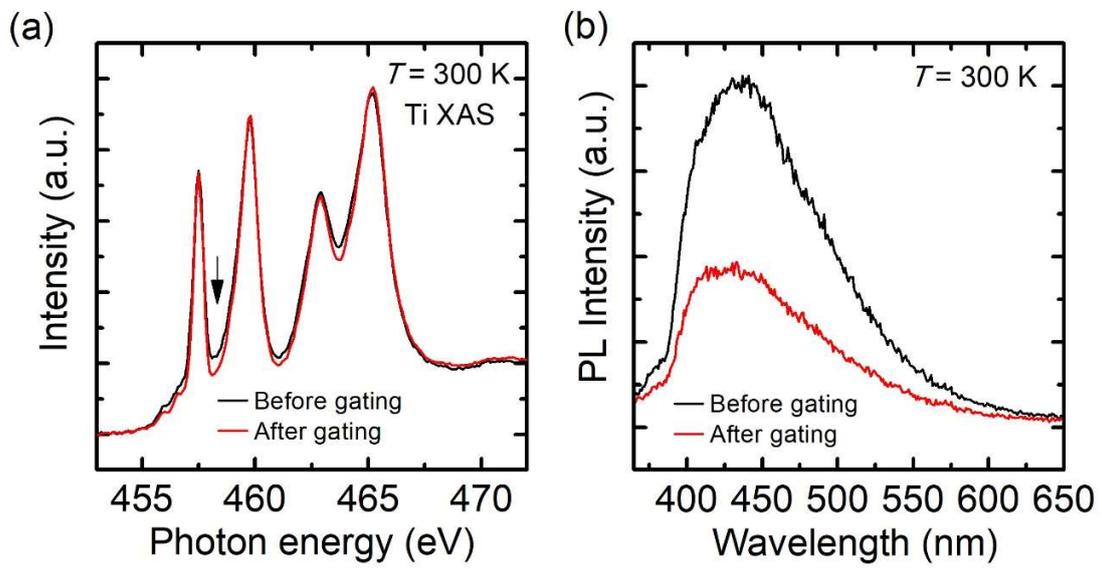

Figure 3

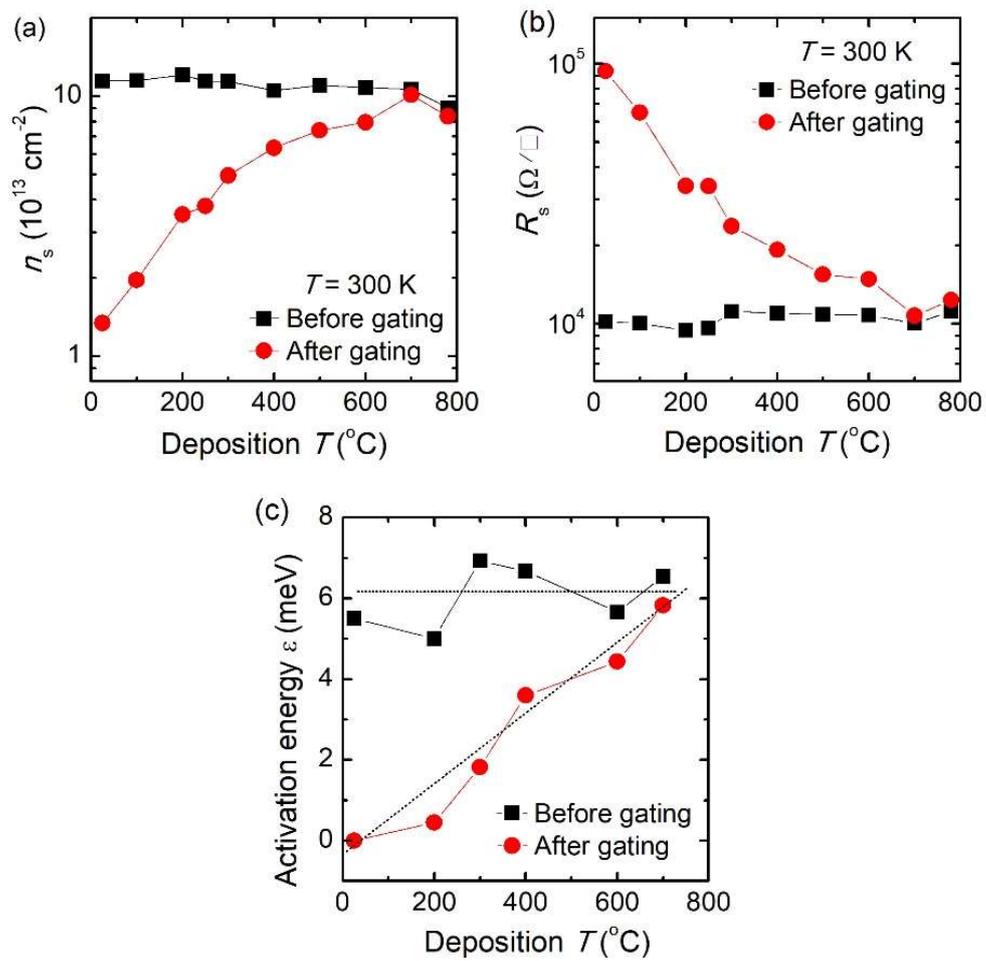

Figure 4

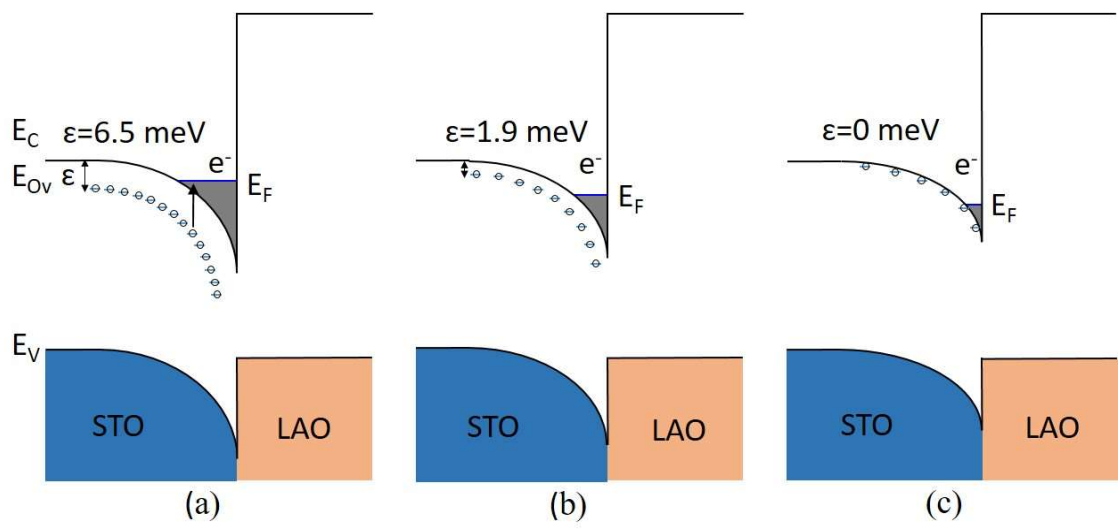

Figure 5

Supplementary materials for
# Oxygen Electromigration and Energy Band Reconstruction Induced by Electrolyte Field Effect at Oxide Interfaces

S. W. Zeng[1,2*], X. M. Yin[2,3,4], T. S. Herng[5], K. Han[1,2], Z. Huang[1,2], L. C. Zhang[1,2], C. J. Li[1,5], W. X. Zhou[1,2], D. Y. Wan[1,2], P. Yang[3], J. Ding[1,5], A. T. S. Wee[2,6], J. M. D. Coey[1,7], T. Venkatesan[1,2,5,8,9], A. Rusydi[1,2,3], A. Ariando[1,2,9*]

[1]NUSNNI-NanoCore, National University of Singapore, Singapore 117411, Singapore
[2]Department of Physics, National University of Singapore, Singapore 117542, Singapore
[3]Singapore Synchrotron Light Source (SSLS), National University of Singapore, 5 Research Link, Singapore 117603, Singapore
[4]SZU-NUS Collaborative Innovation Center for Optoelectronic Science & Technology, Key Laboratory of Optoelectronic Devices and Systems of Ministry of Education and Guangdong Province,
College of Optoelectronic Engineering, Shenzhen University, Shenzhen 518060, China
[5]Department of Materials Science and Engineering, National University of Singapore, Singapore 117576, Singapore
[6]Centre for Advanced 2D Materials and Graphene Research, National University of Singapore, Singapore 117546, Singapore
[7]School of Physics and CRANN, Trinity College, Dublin 2, Ireland
[8]Department of Electrical and Computer Engineering, National University of Singapore, Singapore 117576, Singapore
[9]National University of Singapore Graduate School for Integrative Sciences and Engineering (NGS), 28 Medical Drive, Singapore 117456, Singapore
*To whom correspondence should be addressed.
E-mail: phyzen@nus.edu.sg, ariando@nus.edu.sg

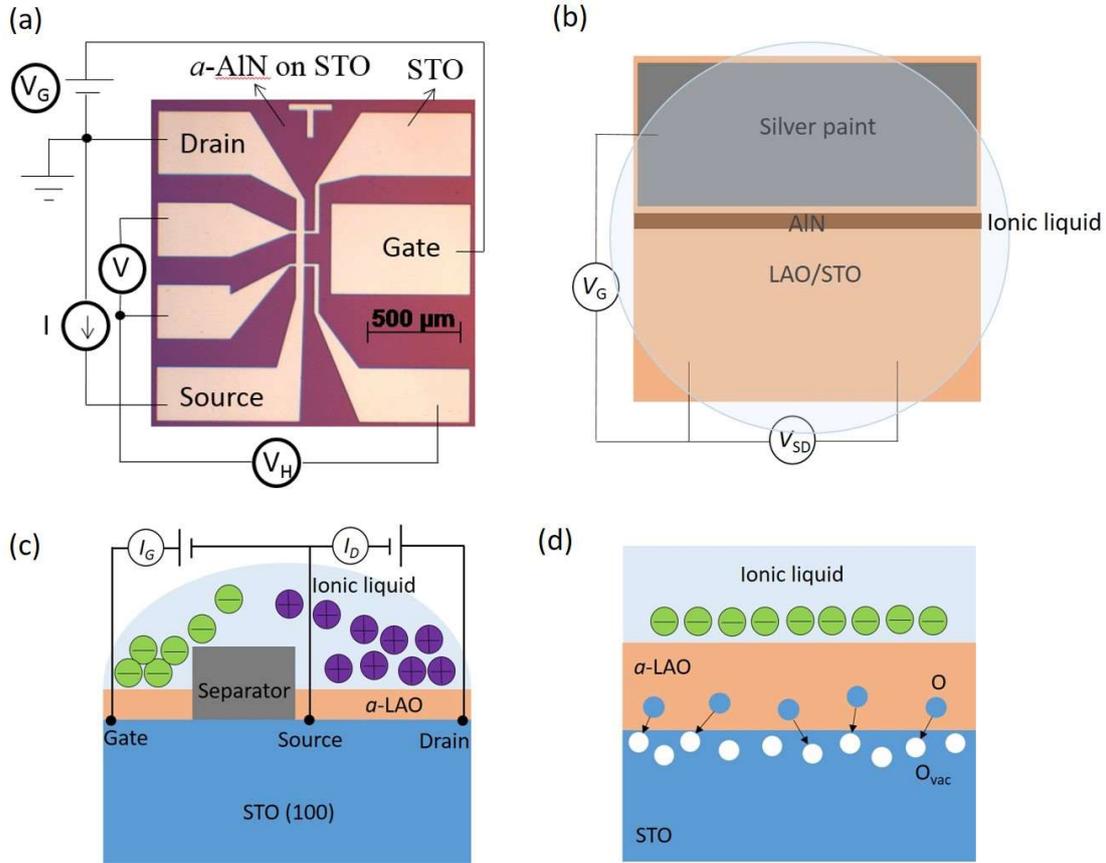

Fig. S1. (a) Optical micrograph of a device pattern and the measurement circuit. The width of the Hall bar is 50 μm and the distance between two voltage probes is 160 μm. The patterning was done by using conventional photolithography and insulating amorphous AlN (*a*-AlN) films were deposited as a hard mask. After patterning, the substrates were annealed in a tube furnace in air at 500 °C for 2 hour to remove the oxygen vacancies which were created during the PLD deposition of *a*-AlN. Then, the patterned substrates were set into PLD chamber for the deposition of LAO. After LAO growth, we did not carry out any further processing step in order to avoid exposure of the samples to chemicals and water which could degrade the quality of surface and interface. The gate electrode was formed by covering silver paint on the lateral LAO gate pad. The wire connection for transport measurement was done by Al ultrasonic wire bonding. (b) Schematic of large-area samples without pattern for PL and XAS measurements. A slim layer of *a*-AlN was deposited on the sample surface with a size of 5 × 5 mm$^2$, and the sample is separated into two roughly equal areas. When gating is required, a droplet of silver paint was put on one half of the sample as the gate electrode, and the IL covered both the silver paint and the other half. The PL and XAS measurements were performed on the half without silver paint for the result after gating, and on the half covered by silver paint for the result before gating. Moreover, we also broken the sample into two parts before performing any measurement, one part was used for performing gating and the other part was kept in pristine state. The measurements on both parts before and after gating were together conducted at the same time, in order to avoid the effect of variation of light intensity. When the measurements were performed, the ILs and silver paint were removed and the samples were cleaned by acetone. (c) Schematic of the operation of the $a$-LAO/STO EDLT. When a $V_G$ is applied, mobile ions accumulate on the sample surface causing a change of charge carriers at the interface. (d) Schematic of oxygen filling on the STO surface when anions accumulate on the sample surface at negative $V_G$.

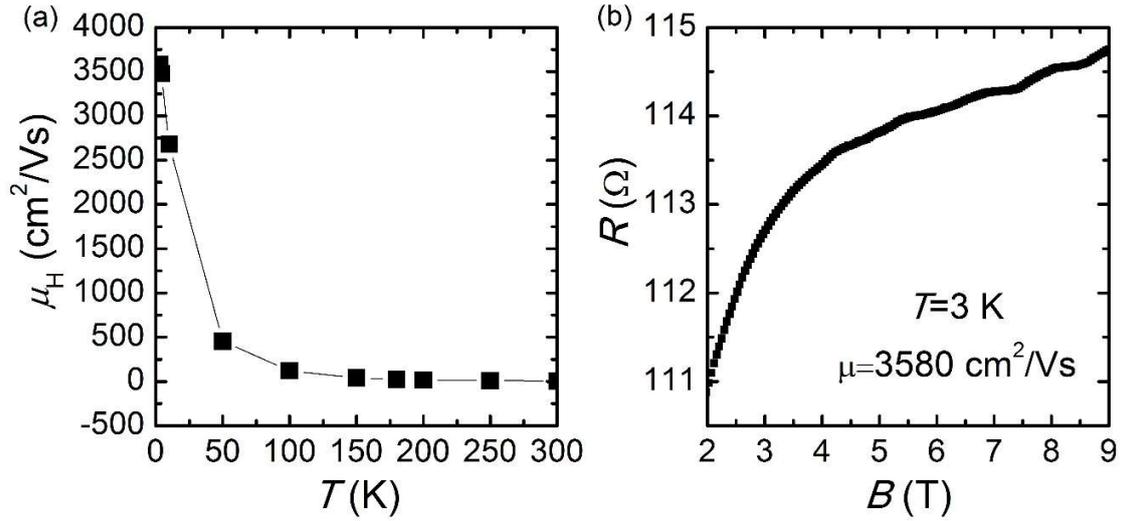

Fig. S2. (a) The mobility $\mu_H$ as a function of temperature for 15-nm $a$-LAO/STO interface after IL gating. The highest mobility is ~3580 cm$^2$/Vs at 3 K. (b) The resistance as a function of magnetic field $B$ at 3 K. Due to the high mobility, the Shubnikov−de Haas (SdH) oscillation of the magnetoresistance is observed for magnetic field above 3.5 T.

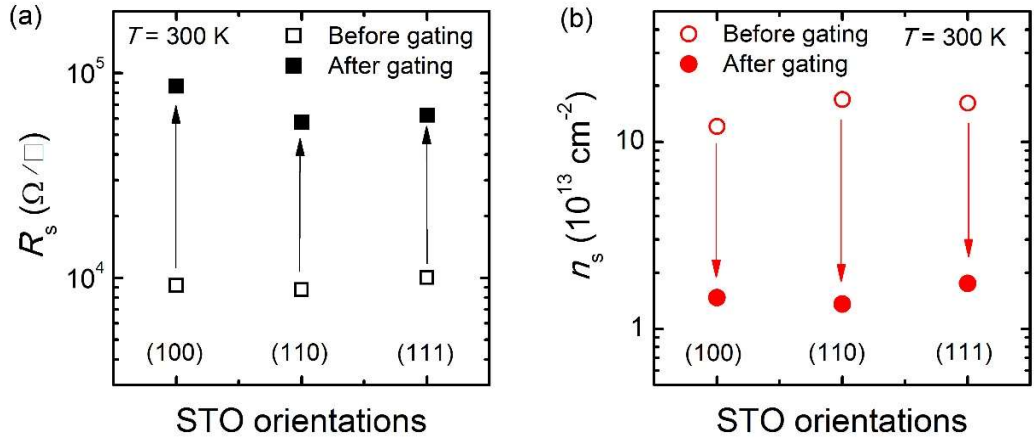

Fig. S3. (a) Sheet resistance $R_s$ and (b) carrier density $n_s$ at $T = 300$ K before and after gating for 2.5-nm $a$-LAO/STO interfaces with different STO orientations of (100), (110) and (111). The arrows denote the change of samples before gating to after gating.

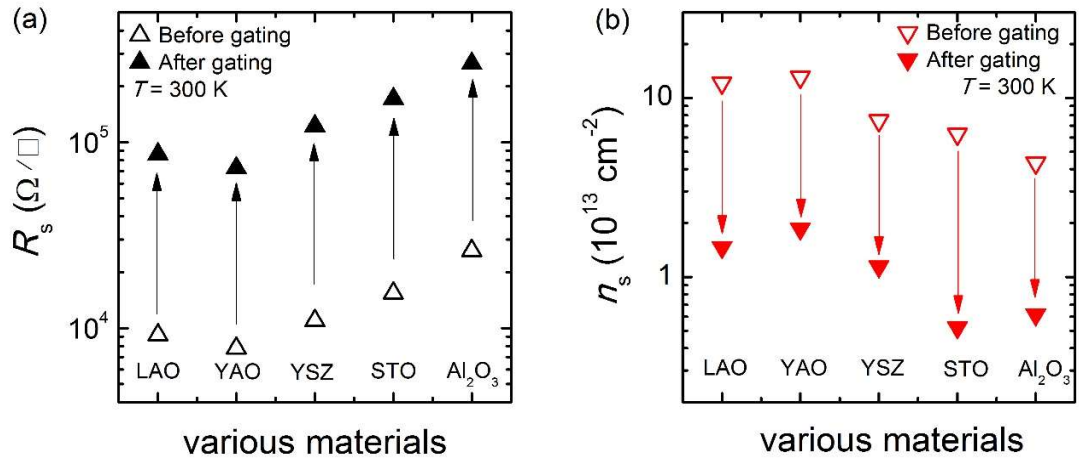

Fig. S4. (a) Sheet resistance $R_s$ and (b) carrier density $n_s$ at $T = 300$ K before and after gating for interfaces between STO (100) substrates and various amorphous materials: LAO, YAlO$_3$ (YAO), yttria-stabilized zirconia (YSZ), STO, Al$_2$O$_3$. The thickness of these amorphous materials is 2.5 nm. The conditions for deposition of these materials are the same to that for deposition of $a$-LAO, as presented in the main text. The targets of LAO, YAO, YSZ and STO are single crystals and the target of Al$_2$O$_3$ is polycrystal. The arrows denote the change of samples before gating to after gating.

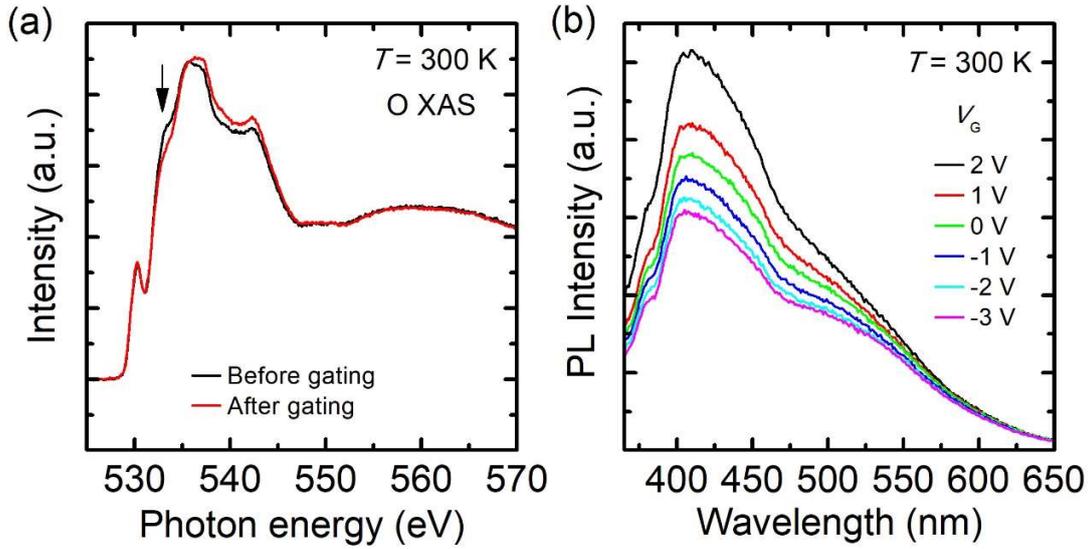

Fig. S5. (a) O $K$-edge XAS spectra for $a$-LAO/STO interfaces before and after IL gating experiments. The black arrow indicates the peak coming from Ti-O $pd$ hybridization with a σ–bonding character. The decrease in this peak intensity for sample after gating indicates the decrease of oxygen vacancies on STO at the interface. (b) In-situ photoluminescence (PL) spectra for another $a$-LAO/STO sample in which the PL measurements were performed during the gating process. The $V_G$ was set to 2 V first and reduced to -3 V at a step of 1 V. One can see that with decreasing $V_G$, the intensity continuously decreases, directly indicating oxygen migration and the resultant decrease of oxygen vacancies in STO in the gating process.

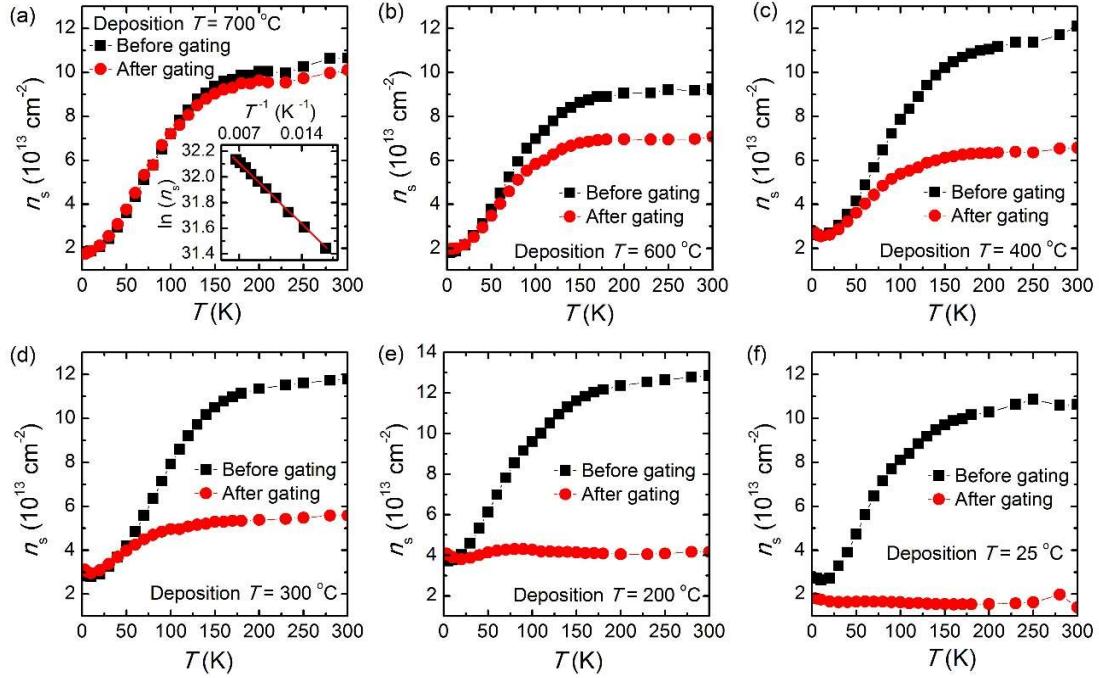

Fig. S6. The carrier density $n_s$ as a function of temperature $T$ for samples before and after gating, with different LAO overlayer deposition $T$. Before gating, $n_s$ of all samples decreases with decreasing $T$ below ~150 K, suggesting carrier freeze out effect. After gating, with decreasing deposition $T$, the samples show less carrier freeze out effect, and at room temperature deposition (25 °C), the $n_s$ is independence of $T$. For the samples which show carrier freeze out effect, the activation energy $\varepsilon$ could be obtained through the fitting using $n_s \propto e^{(-\varepsilon/k_B T)}$, as shown in Fig. 4(c) in the main text. Inset of (a) shows an example of the fitting data for sample after gating in the temperature range between 150 and 50 K.

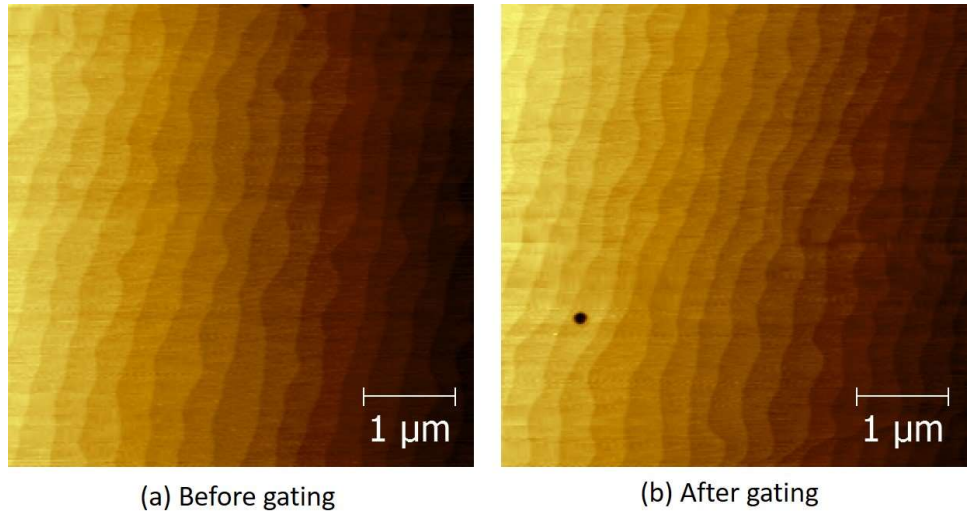

Fig. S7. Surface characterization before and after gating experiments. AFM images of the sample surface (a) before and (b) after gating experiments. The surfaces are significantly similar with clear atomic terraces, indicating that the damage in *a*-LAO/STO EDLTs is negligible.